\documentclass[%
 reprint,
frontmatterverbose, 
showpacs,preprintnumbers,
nofootinbib,
 amsmath,amssymb,
 aps,
floatfix,
]{revtex4-1}

\usepackage{graphicx}
\usepackage{dcolumn}
\usepackage{bm}

\usepackage{amsmath}
\usepackage{slashed}
\usepackage{tikz}
\usepackage{graphicx}
\usepackage{pgfplots}
\usepackage{color}
\usepackage{rotating}
\usepackage{gensymb}
\usepackage{simplewick}
\usepackage{enumerate}
\usetikzlibrary{quotes,angles,snakes}
\usetikzlibrary{trees}
\usetikzlibrary{decorations.pathmorphing}
\usetikzlibrary{decorations.markings}
\usetikzlibrary{arrows,shapes,positioning}
\usetikzlibrary{decorations.markings}
\tikzstyle arrowstyle=[scale=1]
\tikzstyle directed=[postaction={decorate,decoration={markings,
    mark=at position .65 with {\arrow[arrowstyle]{stealth}}}}]
\tikzstyle reverse directed=[postaction={decorate,decoration={markings,
    mark=at position .65 with {\arrowreversed[arrowstyle]{stealth};}}}]

\newcommand{\RNum}[1]{\uppercase\expandafter{\romannumeral #1\relax}}

\newcommand{\beq}{\begin{equation}}
\newcommand{\eeq}{\end{equation}}
\newcommand{\bea}{\begin{eqnarray}}
\newcommand{\eea}{\end{eqnarray}}
\usepackage{ulem}

\usepackage{xcolor,colortbl}
\definecolor{Blue}{RGB}{140,165,195}
\definecolor{Purple}{RGB}{255,145,145}
\definecolor{bluc}{cmyk}{1,1,0,0.1}
\definecolor{rossoCP3}{cmyk}{0,.88,.77,.40}
\definecolor{rosso}{cmyk}{0,1,1,0.4}
\definecolor{rossos}{cmyk}{0,1,1,0.55}
\definecolor{rossoc}{cmyk}{0,1,1,0.2}
\definecolor{verdes}{cmyk}{0.92,0,0.59,0.4}

\newcommand{\SU}{\,{\rm SU}}

\newcommand{\U}{\,{\rm U}}

\usepackage{color}
\usepackage{ulem}

\begin{document}

\preprint{CP3-Origins-2017-025 DNRF90, CERN-TH-2017-143}

\title{ Asymptotically Safe Standard Model  via  Vector-Like Fermions }
\author{R.B.~Mann$^{1}$}
\author{J.R.~Meffe$^{1,2}$}
\author{F.~Sannino$^{2,3,4}$}
\author{T.G.~Steele$^{5}$}
\author{Z.W. Wang$^{1,2}$}
\author{C. Zhang$^{6}$}

\affiliation{$^{1}$Department of Physics, University of Waterloo, Waterloo, On N2L 3G1, Canada\\
$^{2}$$\rm{CP}^3$-Origins, University of Southern Denmark, Campusvej 55
5230 Odense M, Denmark \\
$^{3}$Danish IAS, University of Southern Denmark, Denmark\\
$^{4}$CERN, Theoretical Physics Department, Switzerland, Geneva\\
$^{5}$Department of Physics and Engineering Physics, University of Saskatchewan, Saskatoon, SK, S7N 5E2, Canada\\
$^{6}$Department of Physics, University of Toronto, Toronto, Ontario, Canada  M5S1A7}

\begin{abstract}
We construct asymptotically safe extensions of the Standard Model by adding gauged vector-like fermions. Using large number-of-flavour techniques we argue that all gauge couplings, including the hypercharge and, under certain conditions, the Higgs coupling can achieve an interacting ultraviolet fixed point. 
\end{abstract}
\maketitle

 Although the Standard Model (SM) of particle interactions is an extremely successful theory of nature, it is an effective theory but not a fundamental one. Following Wilson \cite{Wilson:1971bg,Wilson:1971dh}, a theory is fundamental if it features an ultraviolet fixed point. The latter can be either non-interacting (asymptotic freedom) \cite{Gross:1973ju,Cheng:1973nv,Callaway:1988ya,Giudice:2014tma,Holdom:2014hla,Pica:2016krb,Molgaard:2016bqf,Gies:2016kkk,Einhorn:2017jbs,Hansen:2017pwe} or interacting (asymptotically safe) \cite{Litim:2014uca,Litim:2015iea,Esbensen:2015cjw} or mixed \cite{Esbensen:2015cjw,Molgaard:2016bqf,Pelaggi:2017wzr,Abel:2017ujy}.   Except for the non-abelian gauge couplings none of the remaining SM couplings features an ultraviolet fixed point.   
  
%
Here we extend the idea of a safe QCD scenario in \cite{Sannino:2015sel} to the entire SM. We argue that an asymptotically safe completion of the SM can be realized via new vector-like fermions\footnote{ An interesting complementary approach  appeared in  \cite{Bond:2017wut}. Here the authors add new fermions in higher dimensional representations of the SM gauge groups, hoping for a (quasi) perturbative UV fixed point. 
The models were unable to lead to a safe hypercharge and Higgs self-coupling.}. Our work relies on the limit of a large number of fermion matter fields, which allows us to perform a $1/N_F$ expansion  \cite{Holdom:2010qs,Pica:2010mt}. Here the relevant class of diagrams can be summed up to arbitrary loop order, leading to an UV interacting fixed point for the (non) abelian interactions of the SM. Thus,  we go beyond the cornerstone work of \cite{Litim:2014uca} where UV safety is realized in the Veneziano-Witten limit by requiring both $N_c$ and $N_F$ to go to infinity with their ratio fixed, and adjusting it close to the value for which asymptotic freedom is lost.

Depending on how these new vector-like fermions obtain their masses, we can either introduce new scalars  that  generate fermion masses through  new Yukawa operators or, 
simply introduce explicit vector-like mass operators. In the following, we focus on the latter most economical case and explore the following three distinct SM  $\SU(3)\times{\SU_L(2)}\times\U(1)$ charge assignments and multiplicity: 
\begin{itemize}
\item [i)]{ $N_F \,\left(3,2,1/6\right)  $;}
\item [ii)]{    $N_{F3}\, \left(3,1,0\right)  \oplus \,N_{F2} \, \left(1,2,1/2\right) $; }
\item [iii)]   $N_{F3}\,\left(3,1,0\right)  \oplus  N_{F2}\, \left(1,3,0\right)  \oplus  N_{F1}\, \left(1,1,1\right)  $. 
\end{itemize}
To the above one needs to add, for each model, the associated right charge-conjugated fermions.   The above models are to be viewed as templates  that allow us to exemplify our novel approach in the search of an asymptotically safe extension of the SM. The basic criterion is that  different fermions should have the same charge if it is non-zero; otherwise the summation technique fails (see Eq.~\eqref{higher order contribution} and the corresponding discussion). In fact we have checked that other models
(e.g. $N_{F3}\left(3,1,2/3\right)  \oplus N_{F2}\left(1,3,0\right)$
featuring new top primes)  lead to similar results as the ones used here\footnote{ Following our innovative approach, a recent follow-up  paper appeared \cite{Pelaggi:2017abg}, in which the set $N_{F3}$ is abandoned. Here QCD remains asymptotically free while the rest of the SM gauge couplings are still safe.}.  Finally, we neglect (in model (i)) possible mixing among the new vector-like fermions and SM quarks. 

 We start by considering the RG equations describing the gauge-Yukawa-quartic to two loop order including vector-like fermions. We have checked that our results agree  for the SM case with the ones in \cite{Buttazzo:2013uya,Antipin:2013sga}. We used  \cite{Jones:1981we,Machacek:1983tz}  for the vector-like fermions contributions to gauge couplings and \cite{Chetyrkin:2013wya,Machacek:1984zw} for the contributions to the Higgs quartic. The associated beta functions read:
  \begin{equation}
\begin{split} 
&\beta_1=\frac{d\alpha_1}{dt}=\left(b_1+ c_1\alpha_1+d_1\alpha_3+e_1\alpha_2-\frac{17}{3}\alpha_{y_t}\right)\alpha_1^2\\
&\beta_2=\frac{d\alpha_2}{dt}=\left(-b_2+c_2\alpha_2+d_2\alpha_3+e_2\alpha_1-3\alpha_{y_t}\right)\alpha_2^2\\
&\beta_3=\frac{d\alpha_3}{dt}=\left(-b_3+c_3\alpha_3+d_3\alpha_2+e_3\alpha_1-4\alpha_{y_t} \right)\alpha_3^2\\
&\beta_{y_t}=\frac{d\alpha_{y_t}}{dt}=\left(9\alpha_{y_t}-\frac{9}{2}\alpha_2-16\alpha_3-\frac{17}{6}\alpha_1\right)\alpha_{y_t}+\beta_{y_t}^{\rm{2loop}} \\ 
&\beta_{\alpha_{h}}^{\rm{1loop}}=\frac{d\alpha_{h}}{dt}=\frac{3}{8} \left(\alpha _1^2+3 \alpha _2^2+2 \alpha _1 \left(\alpha _2-4\alpha_h\right)+\right.\\
&\qquad\qquad~~~~~\,+\left.64 \alpha_h^2-24 \alpha _2\alpha_h+32 \alpha _h \alpha _{y_t}-16 \alpha _{y_t}^2\right)\\\nonumber
\end{split}
\end{equation}
\begin{equation}
\begin{split}
&\beta_{\alpha_{h}}^{\rm{2loop}}=\frac{1}{6} \left(-4D_{R_3}S_2\left(R_2\right) \alpha _2^2 {N_{F2}}\color{black}\left(2 \alpha _1+6 \alpha _2-15 \alpha_h\right)\right.\\
&-4D_{R_3}D_{R_2} \alpha _1^2 Y^2{N_{F1}}\color{black}\left.\left(2 \alpha _1+2 \alpha _2-5 \alpha _h\right)\right)+\beta_{\alpha_{h\,\rm{2loop}}}^{\rm{SM}}\\
&\beta_{\alpha_{h\,\rm{2loop}}}^{\rm{SM}}=\frac{1}{48} \left(-379 \alpha _1^3-559 \alpha _2 \alpha _1^2-289 \alpha _2^2 \alpha _1+915 \alpha _2^3\right)\\
&\qquad\quad~+\frac{1}{48} \left(1258 \alpha _1^2+468 \alpha _2 \alpha _1-438 \alpha _2^2\right) \alpha _h-312 \alpha _h^3\\
&\qquad\quad~+\frac{1}{48} \left(1728 \alpha _1+5184 \alpha _2\right) \alpha _h^2\label{RG conventional}
\end{split}
\end{equation}
where  $t=\ln\left(\mu/M_Z\right)$ and $\alpha_1$, $\alpha_2$, $\alpha_3$, $\alpha_{y_t}$, $\alpha_h$ are the  $\U(1)$, $\SU(2)$, $\SU(3)$, top-Yukawa and Higgs self-couplings respectively and we have used the normalization 
\begin{equation}
\alpha_i=\frac{g_i^2}{\left(4\pi\right)^2},\quad\alpha_{y_t}=\frac{y_t^2}{\left(4\pi\right)^2},\quad\alpha_h=\frac{\lambda_h}{\left(4\pi\right)^2}\,.
\end{equation}
$\beta_{\alpha_{h\,\rm{2loop}}}^{\rm{SM}}$ and $\beta_{y_t}^{\rm{2loop}}$ represent two loop SM contributions to the RG functions of $\alpha_h$ and $\alpha_{y_t}$, which are not shown explicitly. $D_{R_2}, D_{R_3}$ represent the dimensions of the representations ($R_2,\,R_3$) under $SU(2)$ and $SU(3)$ while $S_2\left(R_2\right)$ represents the Dynkin index of the representation $R_2$. 
The contributions of the SM chiral fermions are encoded in $b_1, b_2,b_3,c_1,c_2,c_3,d_1,d_2,e_2,e_3$ in Eq.\eqref{explicit coefficients} and can be distinguished from the new vector-like contributions that are all proportionals to a ``$D_R$" coefficient
\begin{equation}
\begin{split}
b_1&= \frac{41}{3}+
{\frac{8}{3}Y^2 N_{\text{F}}D_{R_2}D_{R_3}}\color{black},~~c_1=\frac{199}{9} +\frac{8}{3}Y^4N_FD_{R_2}D_{R_3}\\ 
b_2&= \frac{19}{3}-
{\frac{4N_F}{3}D_{R_3}}\color{black}{,~~c_2=\frac{35}{3}+}
{\frac{49 N_F}{3}D_{R_3}} \\ 
b_3&=14-
{\frac{4 N_F}{3}D_{R_2}}\color{black}{,~~c_3=-52+}
{\frac{76 N_F}{3}D_{R_2}}\\
d_1&=\frac{88}{3}+
{\frac{32}{3}Y^2N_FD_{R_2}D_{R_3}}\color{black},~e_1=9 +6 Y^2 N_{F}D_{R_2}D_{R_3}\\
d_2&=24+\frac{16}{3} N_FD_{R_3}\color{black}{,~~e_2=3+4Y^2 N_{\text{F}}D_{R_3}}\\
d_3&=9+
{3 N_FD_{R_2}},~~e_3=\frac{11}{3}+4N_FY^2D_{R_2}\,,
\label{explicit coefficients}
\end{split}
\end{equation}
where for simplicity, the above explicit coefficients only apply to fundamental representations (models (i) and (ii)); for higher dimension representations the corresponding Casimir invariants and the Dynkin index should be incorporated.

\begin{figure}[htb]
\centering
\includegraphics[width=0.6\columnwidth]{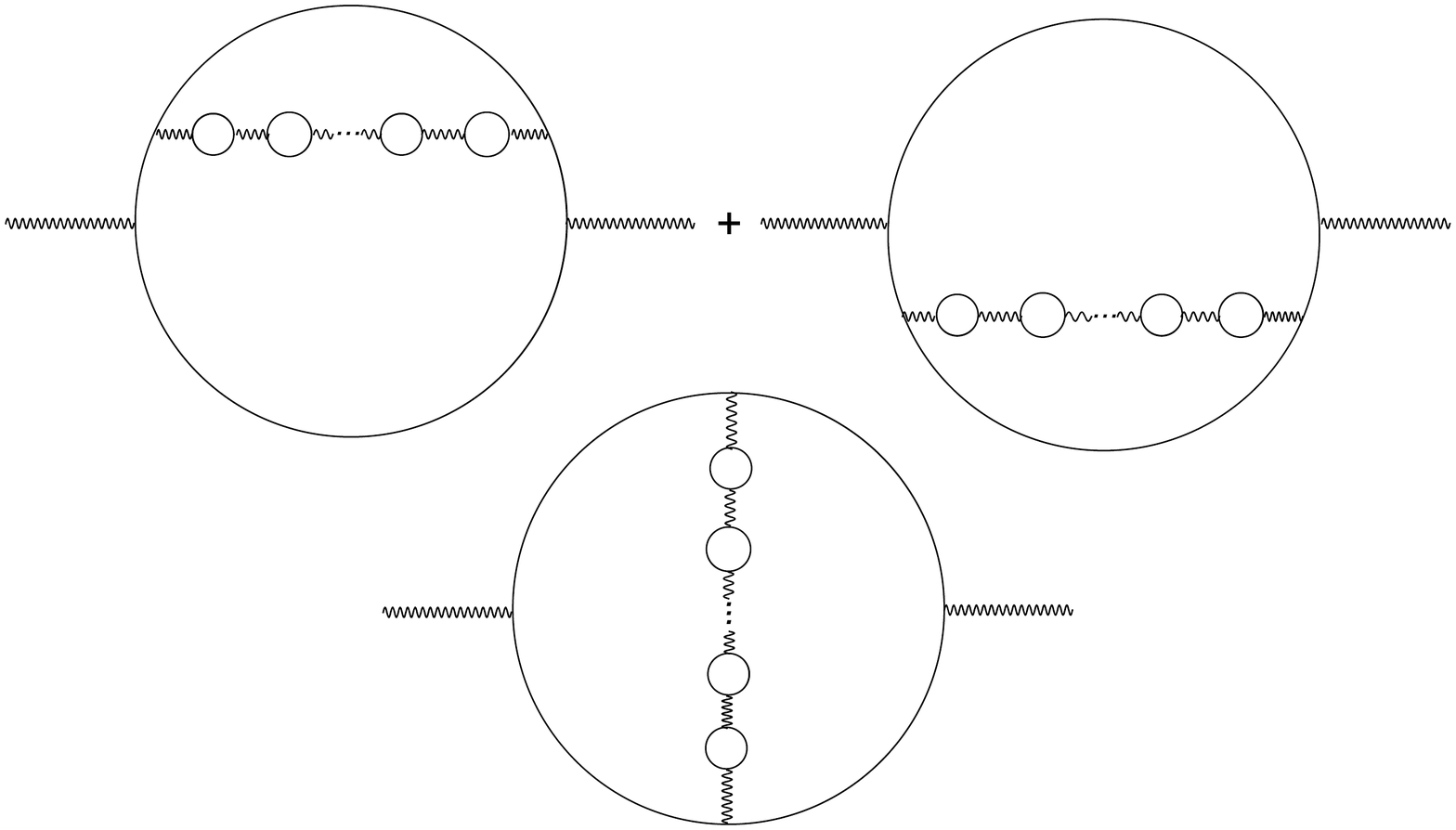}\hspace{0.05\columnwidth}
\caption{Higher order self-energy diagram}
\label{bubble diagram}
\end{figure}

The following diagrams (see Fig.~\ref{bubble diagram}) encode the infinite tower of higher order contributions to the self-energies related to the gauge couplings. These diagrams can be summed up analytically (the abelian and non-abelian cases were first computed respectively in \cite{PalanquesMestre:1983zy} and \cite{Gracey:1996he}).

To the leading $1/N_F$ order, the higher order (ho) contributions to the RG functions of $\beta_2$ and $\beta_3$ are given by \cite{Holdom:2010qs} and have been generalized to the case with any hypercharge $Y$ and semi-simple group ($F_1$ first appeared in \cite{PalanquesMestre:1983zy}):

\begin{equation}
\beta_{\rm{ho}1}=\frac{2A_1\alpha_1}{3}\frac{F_{1}(A_1)}{N_F};~~\beta_{\rm{ho}i}=\frac{2A_i\alpha_i}{3}\frac{H_{1i}(A_i)}{N_F}~~\left(i=2,3\right)\,,\label{higher order contribution}
\end{equation}
where 
\begin{equation}
\begin{split}
A_1&\equiv4\alpha_1N_FY^2D_{R_2}D_{R_3};\quad A_2\equiv 2\alpha_2N_FD_{R_3}\\
A_3&\equiv 2\alpha_3N_FD_{R_2}\\\nonumber
F_1&=\int_0^{A/3}I_1(x)dx\\
H_{1i}&=\frac{-11}{2}N_{ci}+\int_0^{A/3}I_1(x)I_2(x)dx\qquad\left(N_{ci}=2,3\right)\\
I_1(x)&=\frac{\left(1+x\right)\left(2x-1\right)^2\left(2x-3\right)^2\sin\left(\pi x\right)^3}{\left(x-2\right)\pi^3}\\
&\times\left(\Gamma\left(x-1\right)^2\Gamma\left(-2x\right)\right)\\
I_2(x)&=\frac{N_{ci}^2-1}{N_{ci}}+\frac{\left(20-43x+32x^2-14x^3+4x^4\right)}{2\left(2x-1\right)\left(2x-3\right)\left(1-x^2\right)}N_{ci}\,.
\end{split}
\end{equation}
We recall that  the validity of the summation depends on our first criterion which implies that for each gauge group we have only a single $A_i$, constraining the possible vector-like models.  $F_1$ has poles at $A = 15/2 + 3n$ while
 $H_{1i}$ has poles at $A = 3, 15/2,\cdots, 3n+9/2$. In this paper
we concentrate on the   first UV pole branch ($A = 15/2$ for
$F_1$ and $A = 3$ for  $H_{1i}$). Note that the pole structure of  $H_{1i}$ is the 
same for all the non-abelian groups, implying that when $N_F$ is fixed, the non-abelian gauge coupling values will be very close to each other if $D_{R_2}=D_{R_3}$.
The presence of the UV poles at $F_1$ and $H_{1i}$ guarantees the existence of an UV safe fixed point for the gauge couplings.  Note that the  functions $F_1$ and $H_{1i}$ are scheme independent according to \cite{Shrock:2013cca}.  We therefore expect the pole structure and the related UV fixed points to be scheme independent. Physical quantities, such as scaling exponents, were computed in \cite{Litim:2014uca}.
{The $1/N_F^2$ terms are negligible for $N_F$ sufficiently large.  Specifically, as pointed out in \cite{Holdom:2010qs},
for $SU(3)$ one finds that $N_F$ needs to be larger than $32$ while for $U(1)$  one finds $N_F\ge16$.}

Thus the total RG functions for the gauge-Yukawa subsystem can be written as:
\begin{equation}
\begin{split}
\beta_{1tot}&=\beta_1\left(\alpha_{1tot},\alpha_{2tot},\alpha_{3tot},\alpha_{{y_t}tot} \right)+\beta_{\rm{ho}1}\left(\alpha_{1tot}\right)\\
\beta_{3tot}&=\beta_3\left(\alpha_{1tot},\alpha_{2tot},\alpha_{3tot},\alpha_{{y_t}tot} \right)+\beta_{\rm{ho}3}\left(\alpha_{3tot}\right)\\
\beta_{2tot}&=\beta_2\left(\alpha_{1tot},\alpha_{2tot},\alpha_{3tot},\alpha_{{y_t}tot} \right)+\beta_{\rm{ho}2}\left(\alpha_{2tot}\right)\\
\beta_{{y_t}tot}&=\left(9\alpha_{{y_t}tot}-\frac{9}{2}\alpha_{2tot}-16\alpha_{3tot}\right)\alpha_{{y_t}tot}+\beta_{{y_t}tot}^{\rm{2loop}},
\label{tot gauge Yukawa}
\end{split}
\end{equation}
where $\alpha_{itot}$ corresponds to the gauge  couplings including the leading $1/N_F$ contribution to the self-energy diagrams, and $\alpha_{{y_t}tot}$ is the accordingly modified Yukawa coupling. We also avoided the double counting problem due to the simultaneous presence of the  $c_i~\left(i=1,2,3\right)$ terms in Eq.~\eqref{RG conventional} and the leading terms of $\beta_{\rm{ho}2},\,\beta_{\rm{ho}3}$ in Eq.~\eqref{higher order contribution}.  We employ the MS scheme, which is a mass independent RG scheme  allowing us to investigate the running of the couplings independently of the running vector-like masses, except for threshold corrections that can be shown to be controllably small. 

\begin{figure}[htb]
\centering
\includegraphics[width=0.8\columnwidth]{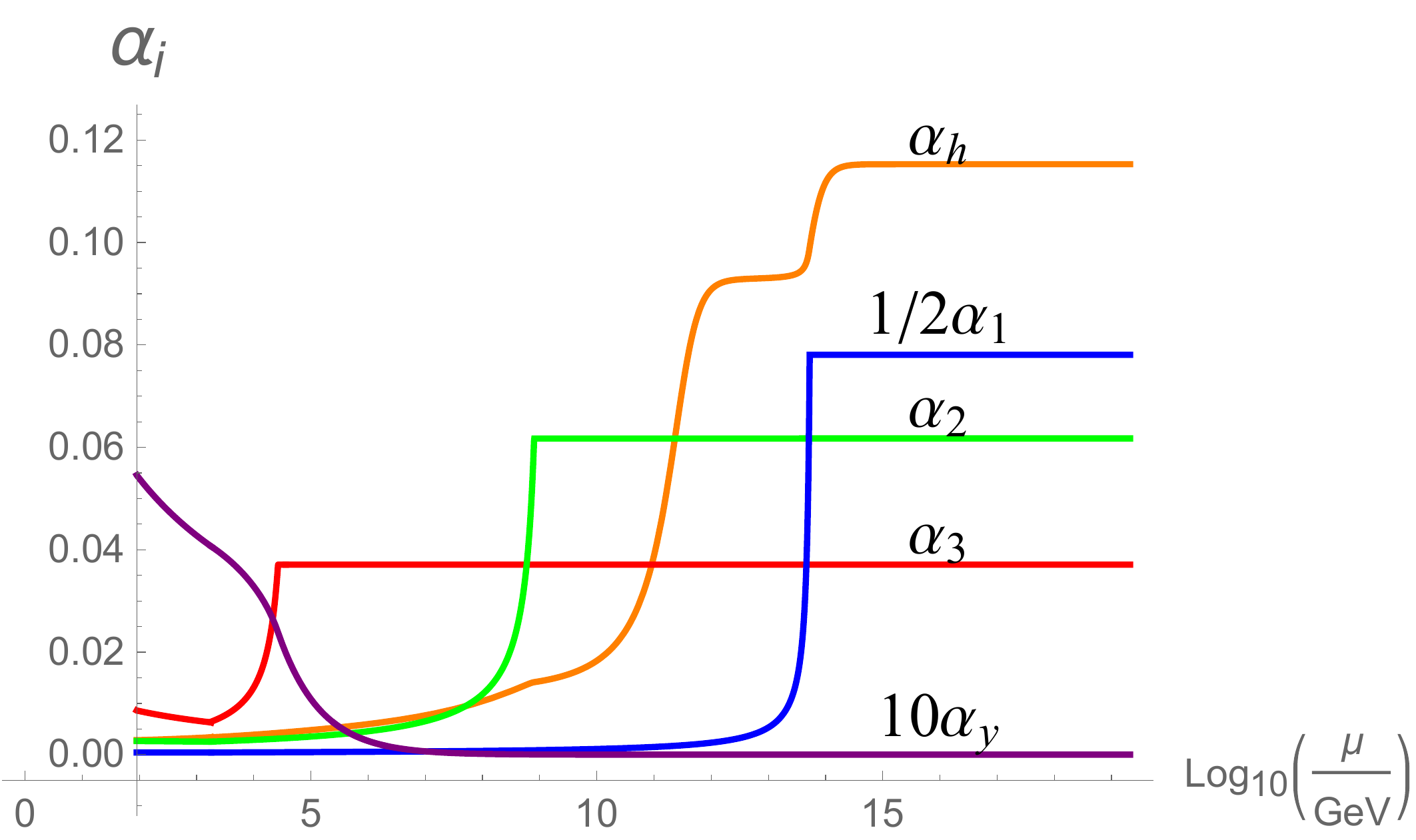}\hspace{0.06\columnwidth}
\caption{ 
Running of the gauge-Yukawa couplings as function of the RG time with $\log_{10}$ base
using model (ii) ($N_{F3}\, \left(3,1,0\right)  \oplus \,N_{F2} \, \left(1,2,1/2\right) $). The blue, green, red and purple curves correspond respectively to the $\U(1)$, $\SU(2)$, $\SU(3)$ gauge  and top Yukawa couplings. The top Yukawa coupling $\alpha_y$ and $U(1)$ gauge coupling $\alpha_1$ have been rescaled by a factor $10$ and $1/2$ respectively to fit all couplings on one figure.  The orange curve depicts the
 two loop level Higgs quartic coupling $\alpha_{h}$ in the same model. Here $N_{F3}=40, N_{F2}=24$ and the initial values of the gauge and Yukawa couplings are chosen to be the SM coupling values at $2\,\rm{TeV}$ while the Higgs quartic coupling is chosen to be $0.0034$.
}
\label{Running Gauge couplings}
\end{figure}

Solving Eqs.~\eqref{tot gauge Yukawa}, we obtain the running coupling solutions depicted in Fig.~\ref{Running Gauge couplings} by the blue, green, red and purple curves, corresponding respectively to the $\U(1)$, $\SU(2)$, $\SU(3)$ gauge couplings and top Yukawa coupling; the orange curve corresponding to the Higgs coupling has not yet been included.  It is clear that all the gauge couplings are UV asymptotically safe while the top Yukawa coupling is asymptotically free. Note that the sub-system {encounters} an interacting UV fixed point at $3.2\times10^{13}\,\rm{GeV}$ which is safely below the Planck scale and so gravity contributions can be safely ignored.  
{For the UV fixed point to exist, the choice of the initial value of the gauge coupling is not crucial since} the only requirement is $\alpha_i\left(t_0\right)<\alpha_i\left(t_*\right),\,\left(i=1,2,3\right)$ where $t_0=\ln\left(\mu_0/M_Z\right)$ is an arbitrary initial scale and $t_*$ is the scale for the UV fixed point. For simplicity, instead of sequentially introducing new vectorlike fermions, we assume they are introduced all at once at a particular scale\footnote{We have checked that our
results change very little if we employ different vector-like fermion masses corresponding to a larger matching scale e.g.~$ m\approx\mu=100\,\rm{TeV}$;  the UV fixed point transition scale  increases accordingly to around $10^{16}\,\rm{GeV}$.}   near their MS-scheme mass $m(m)=m$ ($m\approx\mu=2\,\rm{TeV}$ (or $t=3$) in Fig.~\ref{Running 
Gauge couplings}).
Note that a too small $N_F$ will fail the $1/N_F$ expansion.
To produce Fig.~\ref{Running Gauge couplings}, we have used model (ii)
with $N_{F3}=40, N_{F2}=24$ with the initial values of the gauge and Yukawa couplings chosen to be the SM coupling values at $2\,\rm{TeV}$ corresponding to $t_0=3$: 
\begin{align}
&\alpha _3(t_0)=0.00661\quad \alpha _2(t_0)=0.00256 \nonumber\\
&\alpha _1(t_0)=0.00084\quad \alpha _y(t_0)=0.00403 
\label{SMvalues}
\end{align}
{We emphasize that the basic features of the gauge and Yukawa curves in Fig.~\ref{Running Gauge couplings} are generic and not limited only to model (ii). Figures similar to Fig.~\ref{Running Gauge couplings} result for all} three vector-like fermion models (i,\,ii,\,iii).

\begin{figure}[htb]
\centering
\includegraphics[width=0.7\columnwidth]{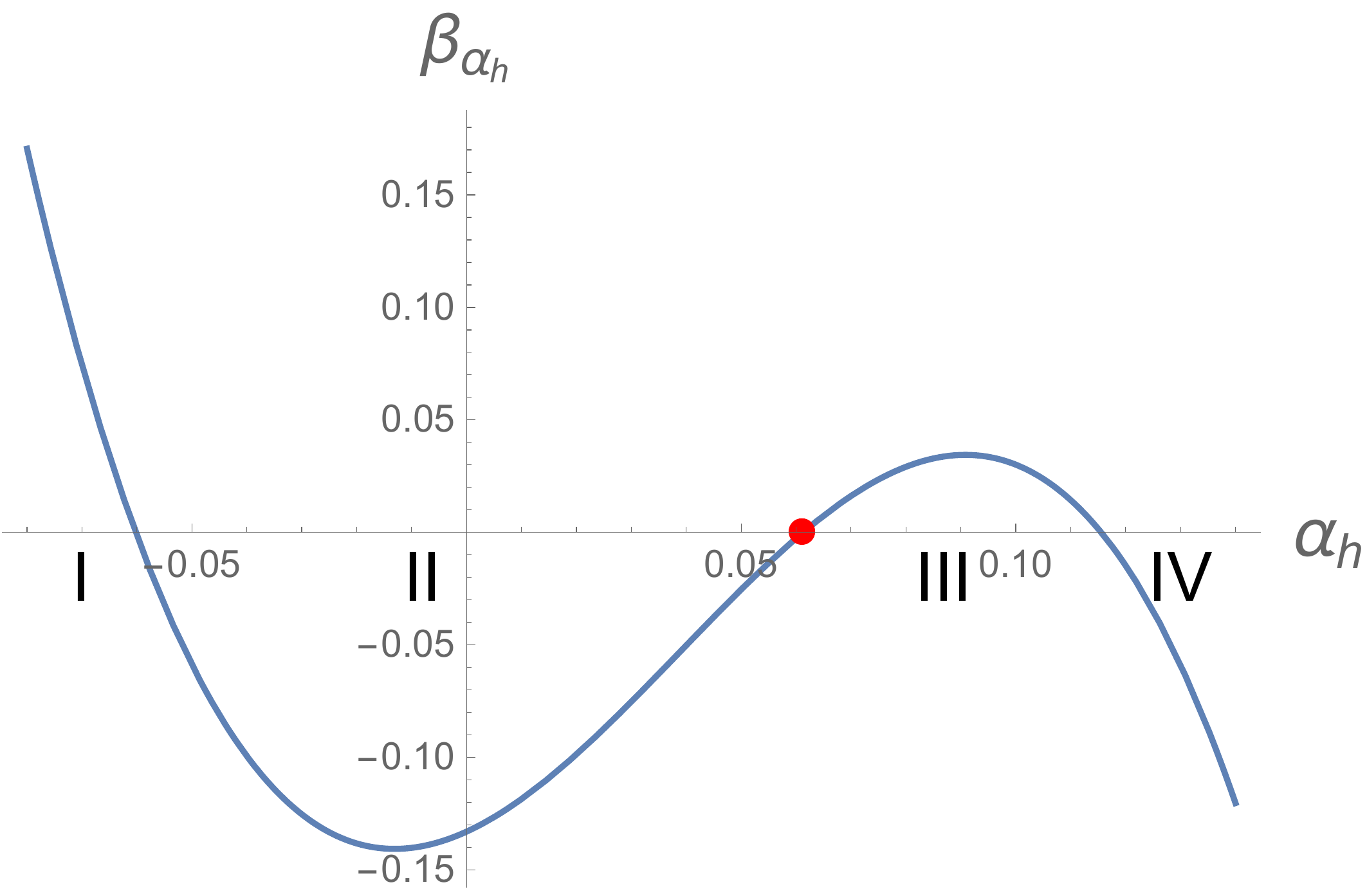}\hspace{0.05\columnwidth}
\caption{This figure shows $\beta_{\alpha_h}$ with $\alpha_h$ with the values of the gauge and Yukawa couplings at the fixed point and $N_{F3}=40, N_{F2}=24$. There exists four different kinds of phases denoted as $I,II,III,IV$ dependent on the initial value of $\alpha_h$. The red point denotes the ultraviolet critical value of $\alpha_h$ which determines whether we could have a UV safe fixed point with positive or negative $\alpha_h$ value.}
\label{beta_Higgs_Quartic_Coupling_2_loop}
\end{figure}

We next consider the Higgs quartic coupling whose beta function to two loop order is given in Eq.~\eqref{RG conventional}. We first plot $\beta_{\alpha_h}$ as a function of $\alpha_h$  for model (ii) with the values of the gauge and Yukawa couplings at the fixed point and $N_{F3}=40, N_{F2}=24$. 
 Fig.~\ref{beta_Higgs_Quartic_Coupling_2_loop} shows that there exist four different regions denoted as $I,II,III,IV$. {Depending on the choice of the initial value of $\alpha_h$, the Higgs self-coupling can be in any of these distinct phases.} Because we are searching for asymptotic safety we are only interested in phase $III$. To guide the reader we mark with a red dot in Fig.~\ref{beta_Higgs_Quartic_Coupling_2_loop}  the ultraviolet critical value\footnote{We distinguish the ultraviolet critical value with the initial critical value of $\alpha_h$, discussed later.  The former quantity is scale dependent; thus the ultraviolet critical $\alpha_h$ is at a scale close to the UV fixed point. The latter quantity is an IR quantity, above which the Higgs self-coupling flows to an UV fixed point; we shall
 take this  inital critical $\alpha_h$  to be at $2\,\rm{TeV}$.} of $\alpha_h$. The plot shows that for the Higgs self-coupling  to be asymptotically safe it must run towards the ultraviolet to values within region $III$, where the other couplings have already reached their fixed point values.  If, however, the dynamics is such that it will run towards ultraviolet values immediately below the critical one the ultimate fate, dictated by phase $II$, is vacuum instability.
 
Fig.~\ref{beta_Higgs_Quartic_Coupling_2_loop} also provides a few insights for  constraining  viable vector-like fermion models. The expression of $\beta_{\alpha_{h}}^{\rm{2loop}}$ in Eq.~\eqref{RG conventional} shows the new vector-like fermions will only provide negative contributions to $\beta_{\alpha_{h}}^{\rm{2loop}}$ when $N_{F2}$ is order of 
10.
{In conjunction} with Fig.~\ref{beta_Higgs_Quartic_Coupling_2_loop}, we expect that the smaller the negative contribution of these new vector-like fermions, the smaller the critical value of $\alpha_h$, and the easier to enter phase $III$. Actually, we find that the pure SM RG function of $\alpha_h$ (without new vector-like fermion contributions to $\beta_{\alpha_h}$ only) provides the smallest critical value of $\alpha_h$, {commensurate with} the above expectation. {Alternatively,}
 if these negative contributions are too large, the cubic curve of $\beta_{\alpha_h}$ will never intersect {the $\alpha_h$ axis} and we will never achieve an asymptotically safe solution (only two phases remain in this limiting case). We learn that the smaller the hypercharge and dimension of the representation, 
the smaller will be the critical value of $\alpha_h$ (making it easier to realize asymptotic safety for the Higgs quartic). Following this criterion, model (ii) 
  should have the smallest critical value of $\alpha_h$. 

 We obtain the same results for the gauge and Yukawa couplings as before, taking their initial values to be the SM ones at $2\,\rm{TeV}$
as in \eqref{SMvalues}. We find that to obtain an asymptotically safe solution for  $\alpha_h$ we must choose its initial value to be (at least)
$\alpha_h(t=3) = 0.0034$, about six times the SM value at that scale. For the SM initial value $\alpha_h(t_0)=0.00054$ the theory achieves the negative value  $\alpha_h=-0.06$ at the UV fixed point, yielding an unstable vacuum. The results for model (ii) (again using $N_{F3}=40, N_{F2}=24$) are 
shown in Fig.~\ref{Running Gauge couplings}, with the Higgs quartic coupling in orange. 

We thus attain UV completion for the whole gauge-Yukawa-Higgs system with gauge and Higgs quartic couplings ($\alpha_1,\alpha_2,\alpha_3,\alpha_h$) asymptotically safe and top Yukawa coupling $\alpha_t$ asymptotically free. The UV fixed point occurs at $3.6\times 10^{14}\,\rm{GeV}$ -- {well below the Planck scale and so} gravity contributions can be safely ignored. {The unique feature in Fig.~\ref{Running Gauge couplings} occurs because when $\alpha_2$ reaches its fixed point value  $\beta_{\alpha_h}$ almost vanishes. However when $\alpha_1$  increases to its final value the almost fixed point in the scalar coupling settles to its true fixed point value.  In addition  this feature, for fixed $N_{F3}=40$,  disappears gradually when increasing $N_{F2}$ from $18$ to $25$. 
This is because the larger $N_{F2}$, the smaller $\alpha_2$ is;   consequently the self-coupling is more sensitive to the change in  $\alpha_1$.}

We have further explored  which regions of parameter space $\left(\alpha_h, N_{F3}, N_{F2}\right)$  can yield asymptotic safety. We find that  $\alpha_h$ {reaches its lowest critical value of $0.0027$ when $N_{F2}=18$ and $32\le N_{F3}\le220$ (insensitive to $N_{F3}$ and the bounds of $N_{F3}$ are discussed below).  This critical $\alpha_h$ value can be further decreased by  considering large $N_F$ of  order a few hundred. Interestingly, there exists an upper value of $N_F$ above which the $A$ in Eq.~\eqref{higher order contribution} goes beyond the first UV pole, moving therefore to the second branch of $F_1$ and $H_1$. Within the first branch, the smallest critical $\alpha_h$ with large $N_F$ occurs for $\alpha_h=0.002$ with $N_{F2}$ near and slightly below the boundary (say $N_{F2}=590$) above which one needs to move to the second branch. The result is insensitive to $N_{F3}$ as well and $32\le N_{F3}\le220$ where the upper bound $N_{F3}=220$ is due to the second branch of $\alpha_3$ while the lower bound $N_{F3}=32$ is to satisfy leading $1/N_F$ expansion.
The UV fixed point occurs below but near the Planck scale. An initial investigation of these other branches suggest that a SM Higgs self-coupling value might be reached, but we leave in-depth investigations for future studies. }

Comparing  models (i) and (ii),  we find that the critical value of $\alpha_h$ is {overall} much higher  for model (i). 
 {However, similar to model (ii), at very large $N_F$ one can decrease $\alpha_h$ below $\alpha_h(t_0)=0.0049$, {corresponding to the lowest critical value one can achieve for small $N_F$}. For example, for an initial value of $\alpha_h=0.0035$ one encounters a UV fixed point provided $N_F\ge105$. It is possible to further decrease $\alpha_h$ with increasing $N_F$. }

For model (iii), we have a similar trend as the previous models. For simplicity, we consider the case where $N_{F1}=N_{F2}$ and note that to achieve  $\alpha_h=0.0035$  (still quite large compared to the SM), one needs $N_{F3}=40$ and $N_{F1}=N_{F2}\ge131$. {Here we find the smallest critical Higgs self-coupling occurs for $\alpha_h=0.00176$ with $N_{F1}=2200, N_{F2}=147, N_{F3}=138$.  These values correspond to the uppermost values allowed by the first branches of the corresponding $F_1$ and $H_1$ functions. This Higgs quartic value is, however, still three times its SM one at $2\,\rm{TeV}$, which is roughly two times the  value at the electroweak scale. We expect that the critical $\alpha_h$ further decreases in the second branch when considering even larger $N_F$.}  We  have checked that our results are stable against the introduction of known higher order terms in $1/N_F$ proportional to the $F_{2\sim4}$ and $H_{2\sim4}$ functions.

Summarising, for all three vector-like-fermion models, with SM gauge and top Yukawa couplings values as initial conditions at IR, we are able to realize UV completion of the gauge-Yukawa subsystem (gauge couplings asymptotically safe and Top Yukawa coupling asymptotically free). Upon including the Higgs quartic coupling, we find that its initial low energy value must attain a certain threshold for a given choice of the number of vector-like fermions. Above this critical value, we attain a  UV asymptotically safe completion, whereas  below this value the system is UV unstable.  For the three vector-like-fermion models we studied, model (ii) possesses the lowest critical value of {$\alpha_h=0.0027$ for a relatively small number of flavours $N_F$. This value is still larger than the (as yet unmeasured) SM Higgs quartic coupling.  If at future colliders the Higgs quartic coupling is found to be $5\--6$ times larger (predicted in some studies without altering the SM RG functions e.g.~\cite{Steele:2012av}), model (ii) could realize asymptotic safety for the whole gauge-Yukawa-Higgs system. Intriguingly an $\alpha_h$ close to the SM value, {say around 2 times at electroweak scale}, can be achieved for very large values of  $N_{F1}$ in model (iii) within the first branch of the $F_1$ and $H_1$ functions. This allows complete asymptotic safety at energies below but near the Planck scale.  

Our results pave the way to  new approaches for  making the SM fully asymptotically safe\footnote{Indeed, building on the present approach in  \cite{Pelaggi:2017abg} it has been shown that one can construct related asymptotically safe SM extensions in which the Higgs quartic coupling matches the SM value. In \cite{Abel:2017rwl} instead, asymptotic safety is achieved via dynamical symmetry breaking of a calculable UV fixed point. }. 

} 
  
\begin{acknowledgments}
T.G.S  and R.B.M are grateful for financial support from the Natural Sciences and Engineering Research Council of Canada (NSERC).  Z.W.~Wang and C.~Zhang thanks Bob Holdom and Jing Ren for very helpful suggestions. F.S. thanks Steven Abel and Alessandro Strumia for insightful discussions. The work is partially supported by the Danish National Research Foundation under the grant DNRF:90. 
\end{acknowledgments}
 

\end{document}